\documentclass[prb,preprint,eqsecnum]{revtex4}%
\usepackage{amsmath}
\usepackage{graphicx}
\usepackage{amssymb}
\usepackage{graphicx}
\usepackage{amsfonts}%
\begin{document}
\title{Effective way to sum over long range Coulomb potentials in two and three dimensions}
\author{Sandeep Tyagi}
\email{satst27@pitt.edu}
\affiliation{Department of Physics and Astronomy, University of Pittsburgh, Pittsburgh,
Pennsylvania 15260}

\begin{abstract}
I propose a method to calculate  logarithmic interaction  in two
dimensions and coulomb interaction  in three dimensions under periodic
boundary conditions.  This paper considers  the case of  a rectangular
cell   in  two   dimensions  and   an  orthorhombic   cell   in  three
dimensions.  Unlike  the  Ewald   method  \cite{ewald},  there  is  no
parameter to be  optimized, nor does it involve  error functions, thus
leading to the  accuracy obtained. This method is  similar in approach
to  that of Sperb\cite{sperb2}  , but  the derivation  is considerably
simpler and physically appealing.  An important aspect of the proposed
method is the faster convergence  of the Green function for a
particular case as compared to Sperb's work.  
The convergence  of the sums  for the most  part of
unit cell is exponential, and hence requires the calculation of only a
few dozen terms. In a very  simple way, we also obtain expressions for
interaction  for systems  with  slab geometries.  Expressions for  the
Madelung constant of CsCl and NaCl are also obtained.

\end{abstract}
\maketitle

\section{Introduction}

In Molecular dynamics (MD) and Monte Carlo (MC) simulations one is required to
calculate the potential energy and forces acting on a particle due to other
particles. Sometimes such forces have a long range interaction. In such
situations, periodic boundary conditions are usually imposed in order to avoid
the boundary effects, which might be especially prominent for small systems
that are usually employed in MD simulations. Under periodic boundary
conditions interaction of a particle with another particle includes the direct
interaction plus an interaction of the first particle with all replicas of
itself as well as all replicas of the second particle. These replicas come
into picture due to the periodic repetitions of a charge under the periodic
boundary conditions. The energy contribution arising from the interaction of a
particle with its own replicas is termed as the self energy. The calculation
of self energy is important in an MC simulation, where size of the box might
change during simulation, such as in isobaric MC. The natural question that
arises is how may one compute the long range interaction of a particle with a
second particle along with all the replicas of the second particle. The self
energy part may then be obtained trivially as well. For eighty years,
researchers have employed the Ewald sum technique to perform such summations.
However, the Ewald sum technique has certain drawbacks. The primary drawback
being, the optimization of a parameter that renders break up of the original
algebraic sum in two parts, one in real space and the other one in Fourier
space. Only when this parameter is chosen properly do the sums in real and
Fourier spaces converge fast. A second problem with the Ewald sum is that even
if one achieves optimal choice of the parameter for breaking up the sum, one
might lose numerical accuracy as the terms in these sums involve error
functions, whose evaluation to high degree of accuracy is difficult. 
In this paper we will consider the logarithmic interaction in two dimensions
and Coulomb interaction in three dimensions. The 2D case has been
satisfactorily dealt with in Ref. \onlinecite{niels}. Thus mainly we
will concentrate on 3D results. The Ewald method is the most widely
used technique for system in 3D. An
alternative technique for summation over long range forces in 3D for a cubic
unit cell was given by Lekner \cite{lekner1}. A tedious method was employed to
obtained the self energy part of the interaction. However, Lekner generalized
his work to an orthorhombic cell \cite{lekner2} and obtained self energies in
a much simpler manner. These recent
methods by Lekner \cite{lekner1} and Sperb \cite{sperb2} are similar in spirit
but their derivation involves complicated algebra. One problem with Lekner's
expressions is that they involve a triple sum. Sperb's \cite{sperb2} 
results are
better in that part of the interaction has only a double sum. 
Nevertheless a triple sum (Eq. 2.4 and 2.7 in
Ref. \onlinecite{sperb2}) is still
employed for the case when both particles are very close to each other.

The technique that we propose is based on a series summation in Fourier space.
Work along these lines has been previously reported in recent papers
\cite{paper1,paper2}, as well as by Harris \textit{et al.} \cite{harris},
Sperb\cite{sperb2}, Crandall \textit{et al.} \cite{crandall} and
Marshall\cite{marshall}. The outline of this paper follows. In section II, we
derive a general formula for dimension $d\geq2$. In section III the formula is
applied to get logarithmic sum in 2D. Section IV describes application of the
general formula to get Coulomb summation $1/r$ for the slab geometry
case as well as for 3D case. Section V considers evaluation of
Madelung constants for CsCl and NaCl. Finally, we discuss our results in
Section VI.

\section{Coulomb sum in $d$ dimension}
An interaction in which satisfies the Poisson equation
in $d$ dimensions will be termed as a Coulomb type potential for that
particular dimension. For example the logarithmic interaction is
a Coulomb type interaction in 2D. In this section we discuss how one
can calculate a pairwise Coulomb interaction between two particles,
separated by a displacement $\mathbf{r}$. For simplicity, we consider
the case of a unit charge situated within an
orthorhombic cell in $d$ dimensions. Let the $d$ sides of the unit cell be
labeled by $l_{1}$, $l_{2}$,$...$,$l_{d}$. The basic unit cell repeats itself
in all $d$ dimensions. The unit charge
interacts with other identical unit charges (for the case of different charges
$q_{1}$ and $q_{2}$ one just gets an extra factor of $q_{1}q_{2}$) situated at
the vertices of the periodic structure. The interaction between
two particles is given by the Green's function in $d$ dimension,
$G(\mathbf{r})$, which satisfies the Poisson equation,
\begin{equation}
\nabla^{2}G(\mathbf{r})=-C_{d}\,\sum_{\mathbf{l}}\delta(\mathbf{r+l}).
\label{2dim}%
\end{equation}
where $\nabla^{2}$ is the Laplacian operator in $d$ dimensions, $\mathbf{l}$
denotes a $d$ dimensional vector, whose components are integer multiples of
$l_{i}$'s and $C_{d}$ is specified by%

\[
C_{d}=\left\{
\begin{array}
[c]{l}%
B_{2}\text{ \ \ \ \ \ \ \ \ \ \ \ \ \ \ \ \ \ \ \ for } d=2,\\
\left(  d-2\right)  B_{d}\text{ \ \ \ \ \ \ \ \ \ for } d>2,
\end{array}
\right.
\]
where $B_{d}$ stands for the coefficient of $(d-1)$ dimensional surface
element in $d$ dimensions,
\begin{equation}
B_{d}=\frac{d\,\,(\pi)^{\frac{d}{2}}}{\Gamma{(}\frac{d}{2}{+1)}}.
\end{equation}
Here $\Gamma(x)$ stands for the gamma function. Thus $B_{2}=2\pi$, $B_{3}=4\pi$
etc. \ We note that coefficients in Eq.(\ref{2dim}) have been chosen such that
$G(\mathbf{r})$ stands for a Coulomb type summation in $d$
dimensions. For example, for $d>2$, we will have $G$ given by,
\begin{equation}
G(x_{1},x_{2},\cdots,x_{d})=\sum_{\{m\}_{d}}\frac{1}{\left\{  (m_{1}%
l_{1}-x_{1})^{2}+(m_{2}l_{2}-x_{2})^{2}+\cdots+(m_{d}l_{d}-x_{d})^{2}\right\}
^{\frac{d-2}{2}}},
\end{equation}
where $\{m\}_{d}$ stands for set of $d$ numbers $m_{1},m_{2},...,m_{d}$. The
summation over each $m_{i}$ runs from $-\infty$ to $+\infty$. The solution
to Eq.(\ref{2dim}) can be easily expressed in Fourier space,
\begin{equation}
G(x_{1},x_{2},\cdots,x_{d})=\frac{C_{d}}{(2\pi)^{2}}\frac{1}{l_{1}l_{2}\cdots
l_{d}}\sum_{\{m\}_{d}}\frac{e^{i2\pi(m_{1}\frac{x_{1}}{l_{1}}+m_{2}\frac
{x_{2}}{l_{2}}+\cdots+m_{d}\frac{x_{d}}{l_{d}})}}{\left\{  \left(  \frac
{m_{1}}{l_{1}}\right)  ^{2}+\left(  \frac{m_{2}}{l_{2}}\right)  ^{2}%
+\cdots+\left(  \frac{m_{d}}{l_{d}}\right)  ^{2}\right\}  }, \label{all}%
\end{equation}
where $0\leq x_i/l_i<1$.
The function $G(x_{1},x_{2},\cdots,x_{d})$, as defined above, diverges since
the term corresponding to all $m$'s being equal to zero blows up. This is
expected since the sum defined in Eq.(\ref{all}) has contribution coming from
an infinite set of identical charges. For the sum in Eq.(\ref{all}) to
make sense we add an infinitesimal term to
the denominator and subtract off a counter term from the whole sum as follows:%

\begin{align}
G(x_{1},x_{2},\cdots,x_{d})  &  =\frac{C_{d}}{(2\pi)^{2}}\frac{1}{l_{1}%
l_{2}\cdots l_{d}}\times\label{g1}\\
&  \lim_{\xi\rightarrow0}\left(  \sum_{\{m\}_{d}}\frac{e^{i2\pi(m_{1}%
\frac{x_{1}}{l_{1}}+m_{2}\frac{x_{2}}{l_{2}}+\cdots+m_{d}\frac{x_{d}}{l_{d}}%
)}}{\left\{  \left(  \frac{m_{1}}{l_{1}}\right)  ^{2}+\left(  \frac{m_{2}%
}{l_{2}}\right)  ^{2}+\cdots+\left(  \frac{m_{d}}{l_{d}}\right)  ^{2}+\left(
\frac{\xi}{l_{d}}\right)  ^{2}\right\}  }-\frac{1}{\left(  \frac{\xi}{l_{d}%
}\right)  ^{2}}\right)  ,\nonumber
\end{align}
where $\xi$ is an infinitesimal parameter which tends to zero. The prescription
employed above amounts to assumption of the presence of a uniform background
charge. For example, let us consider the case of 3D. For every charge,
$q$, one may imagine a uniform distribution of
charge, such that, total charge per unit cell adds up to $-q$. For a charge
neutral periodic system, imposing these kind of background uniform charge
distributions does not matter since total uniform background charge adds up to
zero. However, now a unit charge located within the unit cell at position
$(x_1,x_2,x_3)$ not only interacts with a second charge located at the origin and
its periodic images, but it also interacts with the neutralizing background
charge, compensating the charge of the second particle. This
particular way of introducing the
neutralizing background charge leads to only the intrinsic part\cite{lekner1}
of potential energy. Now, it can be easily verified that Eq.(\ref{g1}) satisfies
the following equation:
\begin{equation}
\nabla^{2}G(\mathbf{r})=-C_{d}\,\sum_{\mathbf{l}}\delta(\mathbf{r+l}%
)+\frac{C_{d}}{l_{1}l_{2}\cdots l_{d}}, \label{22dim}%
\end{equation}
where the last term in Eq.(\ref{22dim}) represents the uniform background
charge. Complete expression for the potential has a term arising from surface
contribution. For the 2D case this turns out to be zero, but for 3D one
obtains a contribution from dipole term\cite{deleeuw}.

Moving further, we can perform one of the $d$ sums in Eq.(\ref{g1})
analytically\cite{gradshteyn},%
\begin{align}
g(x_{d},\{m\},\xi)=  &  \sum_{m_{d}=-\infty}^{\infty}\frac{e^{i2\pi
m_{d}\,\frac{x_{d}}{l_{d}}}}{(m_{d})^{2}+(m_{1}l_{d,1})^{2}+\cdots
+(m_{d-1}l_{d,d-1})^{2}+{\xi}^{2}}\label{all1}\\
&  =\frac{\pi}{\gamma_{d}(\{m\},\xi)}\frac{\cosh\left[  \pi\gamma
_{d}(\{m\},\xi)\left(  1-2\frac{|x_{d}|}{l_{d}}\right)  \right]  }{\sinh\left[
\pi\gamma_{d}(\{m\},\xi)\right]  },\nonumber
\end{align}
where $l_{i,j}$ stands for $l_{i}/l_{j}$ and $\gamma_{d}(\{m\},\xi)$ is
defined as
\begin{equation}
\gamma_{d}(\left\{  {m}\right\}  ,\xi)=\sqrt{(m_{1}l_{d,1})^{2}+\cdots
+(m_{d-1}l_{d,d-1})^{2}+{\xi}^{2}}.
\end{equation}
For convenience we also define%

\begin{equation}
\gamma_{d0}(\left\{  {m}\right\}  ,\xi)=\sqrt{(m_{1}l_{d,1})^{2}%
+\cdots+(m_{d-1}l_{d,d-1})^{2}}. \label{d0}%
\end{equation}
Using Eqs.(\ref{g1}) and (\ref{all1}) one obtains
\begin{align}
G(x_{1},x_{2},\cdots,x_{d})  &  =\frac{C_{d}}{(2\pi)^{2}}\frac{l_{d}}%
{l_{1}l_{2}\cdots l_{d-1}}\label{all3}\\
&  \times\lim_{\xi\rightarrow0}\left(  \sum_{\{m\}_{d-1}}g(x_{d}%
,\{m\},\xi)\prod_{i=1}^{(d-1)}\cos\left(  2\pi m_{i}\,\frac{x_{i}}{l_{i}%
}\right)  -\frac{1}{\xi^{2}}\right)  .\nonumber
\end{align}
In the limit $\xi\rightarrow0$, the term corresponding to all $m_{i}$ being
set to zero in Eq.(\ref{all3}) must be separated out as follows:%

\begin{align}
G(x_{1},x_{2},\cdots,x_{d})  &  =\frac{C_{d}}{(2\pi)^{2}}\frac{l_{d}}%
{l_{1}l_{2}\cdots l_{d-1}}\label{all4}\\
&  \times\left.  \left(  \sum_{\{m\}_{d-1}}^{\prime}g(x_{d},\{m\},\xi
)\prod_{i=1}^{(d-1)}\cos\left(  2\pi m_{i}\,\frac{x_{i}}{l_{i}}\right)
\right)  \right\vert _{\xi=0}\nonumber\\
&  +\frac{C_{d}}{(2\pi)^{2}}\frac{l_{d}}{l_{1}l_{2}\cdots l_{d-1}}\frac
{\pi^{2}}{3}\left\{  1-6\left(  \frac{|x_{d}|}{l_{d}}\right)  +6\left(
\frac{x_{d}}{l_{d}}\right)  ^{2}\right\}  ,\nonumber
\end{align}
where a prime on the summation sign implies that the term corresponding to all
$m_{i}$ being zero is not to be included. In Eq.(\ref{all4}), we separated out
the term corresponding to all $m_{i}$ being set to zero and took the limit
$\xi\rightarrow0$ as follows,%

\begin{equation}
\lim_{\xi\rightarrow0}\left(  \frac{\pi}{\xi}\frac{\cosh\left[  \pi\xi\left(
1-2\frac{|x_{d}|}{l_{d}}\right)  \right]  }{\sinh\left[  \pi\xi\right]  }%
-\frac{1}{\xi^{2}}\right)  =\frac{\pi^{2}}{3}\left\{  1-6\left(  \frac{|x_{d}|%
}{l_{d}}\right)  +6\left(  \frac{x_{d}}{l_{d}}\right)  ^{2}\right\}  .
\end{equation}
Eq.(\ref{all4}) forms the main result derived in this section. It is important
to note that as a result of the symmetry present in the problem, it suffices
to look at only that part of the unit cell which corresponds to 
$0 \leq x_{i}/l_{i} \leq 0.5$ for all $i$'s. Hence, from here on we
will assume $0 \leq x_{i}/l_{i} \leq 0.5$. In the next two sections,
we investigate two important cases corresponding to $d=2$ and $d=3.$

\section{Logarithmic sum in two dimensions}
Energy of $N$ particles contained in a rectangular unit cell with
periodic boundaries and
interacting through a logarithmic potential in 2D can be expressed as\cite{sperb2},
\begin{equation}
E_{\text{total}}^{\text{2d}}=\frac{1}{2}\sum_{i,j;i \neq j}q_{i}q_{j}G_{\text{2d}}(\mathbf{r}_{i}%
-\mathbf{r}_{j})+\sum_{i}q_{i}^{2}G_{\text{self}}^{\text{2d}},\label{tot}%
\end{equation}
where charges are denoted by $q_{i}$ and the position of charges in the unit
cell by $\mathbf{r}_{i}$ where $1\leq i\leq N$. We will obtain expressions for
$G_{\text{2d}}(\mathbf{r})$ and $G_{\text{self}}^{\text{2d}}$ in this section. The pairwise
interaction is given by the Green function $G_{\text{2d}}(\mathbf{r})$ which
satisfies the Poisson equation in 2D,
\begin{equation}
\nabla^{2}G_{\text{2d}}(\mathbf{r})=-2\pi\sum_{\mathbf{l}}\delta
(\mathbf{r+l})+\frac{2\pi}{l_{1}l_{2}}, \label{logarithm1}%
\end{equation}
where the last term on the rhs of Eq.(\ref{logarithm1}) stands for the neutralizing
background charge.
Eq.(\ref{logarithm1}) is a special case of Eq.(\ref{2dim}). We look for a
solution of Eq.(\ref{logarithm1}) with periodic boundary conditions along
$x_{1}$ and $x_{2}$ directions. This solution can be easily obtained
from the general formula, Eq.(\ref{all4}), derived in the previous
section,

\begin{align}
G_{\text{2d}}(x_{1},x_{2})  &  =\frac{1}{2\pi}\frac{l_{2}}{l_{1}}%
\sum_{m^{\prime}}\frac{\pi}{\gamma_{20}(m)}\frac{\cosh\left[  \pi\gamma
_{20}(m)\left(  1-2\frac{|x_{2}|}{l_{2}}\right)  \right]  }{\sinh\left[
\pi\gamma_{20}(m)\right]  }\cos\left(  2\pi m\frac{x_{1}}{l_{1}}\right)
\label{2dg1}\\
&  +\frac{1}{2\pi}\frac{l_{2}}{l_{1}}\frac{\pi^{2}}{3}\left\{  1-6\left(
\frac{|x_{2}|}{l_{2}}\right)  +6\left(  \frac{x_{2}}{l_{2}}\right)
^{2}\right\},\nonumber
\end{align}
where a prime on $m$ implies the term corresponding to $m=0$ is to be
excluded. Without any loss of generality we may assume that sides of the rhombic
cells have been labeled so that $l_1 \leq l_2$. This condition will
make sure that $\gamma_{20}(m)>1$ for all integer values of $m$. Let us now consider
the convergence of the sum in Eq.(\ref{2dg1}). The first part of
Eq.(\ref{2dg1}) converges exponentially, but in some cases
the convergence may be very slow. Specifically, the leading term in
(\ref{2dg1}) decays as $\exp\left(-2 \pi |m| |x_2|/l_1 )\right)$. Thus the
convergence depends on the ratio $x_2/l_1$. We see that one obtains a slow
exponential convergence when $0 \leq x_{2}/l_1 <0.1$. To handle this case properly, we
break the first sum in Eq.(\ref{2dg1}) into two parts by application of a
trigonometric identity,%

\begin{equation}
\frac{\cosh(a-b)}{\sinh\left(  b\right)  }=\frac{\cosh(a)~\exp(-{b})}%
{\sinh(b)}+\exp(-a). \label{break}%
\end{equation}
This leads to the expression:
\begin{align}
&  \frac{1}{2\pi}\sum_{m^{\prime}}\frac{\pi}{\left\vert m\right\vert }%
\frac{\cosh\left[  \pi ml_{2,1}\left(  1-2\frac{|x_{2}|}{l_{2}}\right)  \right]
}{\sinh(\pi\left\vert m\right\vert l_{2,1})}\cos\left(  2\pi m\frac{x_{1}%
}{l_{1}}\right) \label{2dg2}\\
&  =\frac{1}{\pi}\sum_{m=1}^{\infty}\frac{\pi}{m}\frac{\exp\left(
-\pi\left\vert m\right\vert l_{2,1}\right)  \cosh\left[  \pi m~l_{2,1}\left(
2\frac{x_{2}}{l_{2}}\right)  \right]  }{\sinh(\pi m~l_{2,1})}\cos\left(  2\pi
m\frac{x_{1}}{l_{1}}\right) \nonumber\\
&  +\frac{1}{\pi}\sum_{m=1}^{\infty}\frac{\pi}{m}\exp\left(  -2\pi
m\frac{|x_{2}|}{l_{1}}\right)  \cos\left(  2\pi m\frac{x_{1}}{l_{1}}\right)
.\nonumber
\end{align}
We notice that the first part of Eq.(\ref{2dg2}) converges even for
the case when $0 \leq x_{2}/l_1<0.1$. In fact the slowest convergence for the first part
will now occur for the case when $2x_{2}=l_{2}$. But even this "slowest"
convergence amounts to a very rapid exponential convergence of
$\exp\left(-\pi |m|l_2/l_1 )\right)$.
We have yet to
account for the last sum in Eq.(\ref{2dg2}). Using the identity
\begin{align}
\sum_{n=1}^{\infty}\frac{1}{n}\exp(-2n\pi x)  &  \cos(2\pi ny)=\label{iden1}\\
&  -\frac{1}{2}\ln\left[  \cosh(2\pi x)-\cos(2\pi y)\right]  +\pi x-\frac
{\ln\left(  2\right)  }{2}\text{ \ \ \ \ \ \ \ \ \ }x>0,\nonumber
\end{align}
the last part of the sum in Eq.(\ref{2dg2}) may be explicitly evaluated to%

\begin{equation}
 -\frac{1}{2}\ln\left\{  \cosh\left(  2\pi\frac{x_{2}}{l_{1}}\right)
-\cos\left(  2\pi\frac{x_{1}}{l_{1}}\right)  \right\}  
+\pi\frac{|x_{2}|}{l_{1}}-\frac{\ln\left(  2\right)  }{2}.
\end{equation}
Assembling the terms together, we finally obtain the following expression for
the 2D Green function,%

\begin{align}
G_{\text{2d}}(x_{1},x_{2})  &  =\frac{1}{2\pi}\sum_{m^{\prime}}\frac{\pi
}{\left\vert m\right\vert }\frac{\exp\left(  -\pi\left\vert m\right\vert
\frac{l_{2}}{l_{1}}\right)  \cosh\left[ 2 \pi
  m\frac{x_{2}}{l_{1}}\right]}{\sinh\left(  \pi\left\vert
m\right\vert \frac{l_{2}}{l_{1}}\right)  }\cos\left(  2\pi m\frac{x_{1}}%
{l_{1}}\right) \\
&  -\frac{1}{2}\ln\left\{  \cosh\left(  2\pi\frac{x_{2}}{l_{1}}\right)
-\cos\left(  2\pi\frac{x_{1}}{l_{1}}\right)  \right\} \nonumber\\
&  +\frac{\pi l_{2}}{6l_{1}}\left\{  1+6\left(  \frac{x_{2}}{l_{2}}\right)
^{2}\right\}  -\frac{\ln\left(  2\right)  }{2}.\nonumber
\end{align}
Self energy may be easily obtained as%

\begin{align}
G_{\text{self}}^{\text{2d}}  &  =\lim_{\left(  x_{1},x_{2}\right)  \rightarrow\left(
0,0\right)  }\left\{  G_{\text{2d}}(x_{1},x_{2})+\ln\left(  \sqrt{x_{1}^{2}+x_{2}^{2}%
}\right)  \right\} \\
&  =\frac{1}{2\pi}\sum_{m^{\prime}}\frac{\pi}{\left\vert m\right\vert }%
\frac{\exp\left(  -\pi\left\vert m\right\vert \frac{l_{2}}{l_{1}}\right)
}{\sinh\left(  \pi\left\vert m\right\vert \frac{l_{2}}{l_{1}}\right)  }%
-\ln\left( \frac{ 2\pi}{l_1}\right)+\frac{\pi}{6}\frac{l_{2}}{l_{1}}.\nonumber
\end{align}
The results derived here may be trivially generalized to the case of a rhombic
cell, but our concern in this paper has only been with orthorhombic cases. The
results obtained here were numerically checked and found to be in agreement
with those of Gr{\o}nbech-Jensen \cite{niels}.

\section{Coulomb sum in 3D}
Energy of $N$ particles contained in a orthorhombic unit cell with 
periodic boundaries and
interacting through a Coulomb type potential in 3D can be expressed as,
\begin{equation}
E_{\text{total}}^{\text{3d}}=\frac{1}{2}\sum_{i,j;i \neq j}q_{i}q_{j}G_{\text{3d}}(\mathbf{r}_{i}%
-\mathbf{r}_{j})+\sum_{i}q_{i}^{2}G_{\text{self}}^{\text{3d}}+\frac{2\pi}{3}\left(
\sum_{i}q_{i}\mathbf{r}_{i}\right)  ^{2},\label{tot1}%
\end{equation}
where charges are denoted by $q_{i}$ and the position of charges in the unit
cell by $\mathbf{r}_{i}$ and $1\leq i\leq N$. We will obtain expressions for
$G_{\text{3d}}(\mathbf{r})$ and $G_{\text{self}}^{\text{3d}}$ in this section.
The application of Eq.(\ref{all4}) for an orthorhombic cell in 3D leads to%

\begin{align}
G_{\text{3d}}(x_{1},x_{2},x_{3})  &  =\frac{1}{\pi}\frac{l_{3}}{l_{1}l_{2}%
}\sum_{m_{1},m_{2}}^{\prime}\frac{\pi}{\gamma_{30}(\{m\})}\frac{\cosh\left[
\pi\gamma_{30}(\{m\})\left(  1-2\frac{|x_{3}|}{l_{3}}\right)  \right]  }%
{\sinh\left[  \pi\gamma_{30}(\{m\})\right]  }\label{g3}\\
\times &  \prod_{i=1}^{2}\cos\left(  2\pi m_{i}\,\frac{x_{i}}{l_{i}}\right)
+\frac{l_{3}}{l_{1}l_{2}}\frac{\pi}{3}\left\{  1-6\left(  \frac{|x_{3}|}{l_{3}%
}\right)  +6\left(  \frac{x_{3}}{l_{3}}\right)  ^{2}\right\}  ,\nonumber
\end{align}
where%

\begin{equation}
\gamma_{30}(\left\{  {m}\right\}  ,\xi)=\sqrt{(m_{1}l_{3,1})^{2}+(m_{2}%
l_{3,2})^{2}}.
\end{equation}
Without any loss of generality we assume that axis have been labeled such
that
\begin{equation}
l_{3}\geq l_{2}\geq l_{1}. \label{inequal}.%
\end{equation}
The condition in Eq.(\ref{inequal}) makes sure $\gamma_{30}(\{m\})>1$ 
for all sets $\{m\}$. Eq.(\ref{g3}) is one of our main results for 3D
case. We note that the
potential energy obtained consists of only the intrinsic part\cite{lekner1}. A
dipole contribution will have to be included in Eq.(\ref{g3}) to obtain the
real potential energy \cite{lekner1,deleeuw}. This dipole contribution
is represented by the last term on the rhs in Eq.(\ref{tot1}).We notice that the sum in
Eq.(\ref{g3}) converges exponentially. In fact the terms corresponding to
large $|m_{1}|$ and $|m_{2}|$ decay as $\exp\left(  -2\pi x_{3}\sqrt{\left(
m_{1}/l_{1}\right)  ^{2}+\left(  m_{2}/l_{2}\right)  ^{2}}\right)  $, which
with the assumption in Eq.(\ref{inequal}) means that terms decay faster than
$\exp\left(  -2\pi x_{3}\sqrt{\left(  m_{1}/l_{2}\right)  ^{2}+\left(
m_{2}/l_{2}\right)  ^{2}}\right)  $. Thus the convergence depends upon the
ratio $r_{32}=x_{3}/l_{2}$. For $r_{32}>0.1$, the convergence of series in
Eq.(\ref{g3}) is extremely good. However, the convergence slows down
for the case when $r_{32}<0.1$. This problem may be solved as follows. Applying the
identity from Eq.(\ref{break}) again, we break the first sum in Eq.(\ref{g3})
in three parts%

\begin{equation}
G_{\text{3d}}(x_{1},x_{2},x_{3})=G_{\text{ELC}}(x_{1},x_{2},x_{3}%
)+G_{\text{slab}}(x_{1},x_{2},x_{3})+\frac{l_{3}}{l_{1}l_{2}}\frac{\pi}%
{3}\left\{  1+6\left(  \frac{x_{3}}{l_{3}}\right)  ^{2}\right\}
,\label{combine}%
\end{equation}
where%

\begin{align}
G_{\text{ELC}}(x_{1},x_{2},x_{3}) &  =\frac{1}{\pi}\frac{l_{3}}{l_{1}l_{2}%
}\sum_{m_{1},m_{2}}^{\prime}\frac{\pi}{\gamma_{30}(\{m\})}\frac{\exp\left(
-\pi\gamma_{30}(\{m\})\right)  \cosh\left[  \pi\gamma_{30}(\{m\})\left(
2\frac{x_{3}}{l_{3}}\right)  \right]  }{\sinh\left[  \pi\gamma_{30}%
(\{m\})\right]  }\label{elc}\\
&  \times\prod_{i=1}^{2}\cos\left(  2\pi m_{i}\,\frac{x_{i}}{l_{i}}\right)
,\nonumber
\end{align}
and%

\begin{align}
G_{\text{slab}}(x_{1},x_{2},x_{3})  &  =\frac{1}{\pi}\frac{l_{3}}{l_{1}l_{2}%
}\sum_{m_{1},m_{2}}^{\prime}\frac{\pi}{\gamma_{30}(\{m\})}\exp\left(
-2\pi\gamma_{30}(\{m\})\frac{|x_{3}|}{l_{3}}\right) \label{s2}\\
&  \times\prod_{i=1}^{2}\cos\left(  2\pi m_{i}\,\frac{x_{i}}{l_{i}}\right)
-\frac{2\pi}{l_{1}l_{2}}|x_{3}|.\nonumber
\end{align}
We note an important aspect of this break up of the sum in Eq.(\ref{g3}) in
two parts. Eq.(\ref{s2}) is independent of $l_{3}$ as $l_{3}/\gamma
_{30}(\{m\})$ does not depend on $l_{3}.$ In fact the expression in
Eq.(\ref{s2}) is a three dimensional Coulomb sum for a cell which is open along
the $x_{3}$ direction and periodic along $x_{1}$ and $x_{2}$. Thus the sum in
Eq.(\ref{s2}) corresponds to the slab geometry. Note that the suffix ELC
stands for the so called "Electrostatic correction term ", a phrase borrowed
from Ref. \onlinecite{arnold2}. At this point it is worth while to recast the
last term in Eq.(\ref{combine}) in a different form, which will prove
to be useful later
in the discussion. Suppose we have $n$ charges in a charge neutral
unit cell $\sum_{i} q_i=0$. Let us assume that the position of the
$q_i$ is denoted by $(x_{1i},x_{2i}, x_{3i})$. Then the third term in
Eq.(\ref{combine}) will give rise to a term in the total energy. This
term will be given by
\begin{equation}
E_z=\frac{2 \pi}{l_1 l_2 l_3} \left( \frac{1}{2}\sum_{i,j}q_i q_j |x_{3i}-x_{3j}|^2\right),
\end{equation}
which after expanding the argument, and using the charge neutrality condition
gives
\begin{equation} 
E_z= -\frac{2 \pi}{V} M_{3}^2,
\label{dipole}
\end{equation}
where $M_3=\sum_{i} q_i x_{3 i}$ stands for the total dipole moment along the $x_3$ direction. 

Let us now consider the convergence of $G_{\text{ELC}}$ and $G_{\text{slab}}$. The
function $G_{\text{ELC}}$ decays as $\exp\left(  -2\pi\gamma_{30}%
(\{m\})[1-|x_{3}|/l_{3}]\right)  $. Thus we see that $G_{\text{ELC}}$ converges
exponentially fast for $0\leq r_{3}\leq0.5$. In fact the slowest convergence
of $G_{\text{ELC}}$ occurs for the case $r_{3}=0.5,$ but even this slowest
convergence goes as $\exp\left(  -\pi\gamma_{30}(\{m\})\right)  $, which is
extremely fast keeping in mind the inequality of Eq.(\ref{inequal}).

Now we consider the convergence of $G_{\text{slab}}$. The previously mentioned
problem of convergence still persists and $G_{\text{slab}}$ fails to converge
fast when $0 \leq r_{32}<0.1$. So, the next step is to separate out this
diverging behavior towards small value of $r_{32}$. For that purpose we break
the sum over $m_{i}$'s in Eq.(\ref{s2}) as follows:%

\[
\sum_{m_{1},m_{2}}^{\prime}=\sum_{m_{2}^{\prime}}+\sum_{m_{1}^{^{\prime}%
},m_{2}},
\]
where $m_{1}^{\prime}$ implies that the term corresponding to $m_{1}=0$ is not
to be included. Thus we break up $G_{\text{slab}}$ as%

\begin{equation}
G_{\text{slab}}(x_{1},x_{2},x_{3})=G_{1}(x_{2},x_{3})+G_{2}(x_{1},x_{2}%
,x_{3}), \label{g34}%
\end{equation}
where%

\begin{equation}
G_{1}(x_{2},x_{3})=\frac{1}{\pi}\frac{l_{3}}{l_{1}l_{2}}\left\{  \frac
{2}{l_{3,2}}\sum_{m_{2}=1}^{\infty}\frac{\pi}{m_{2}}\exp\left(  -2\pi
m_{2}\frac{|x_{3}|}{l_{2}}\right)  \cos\left(  2\pi m_{2}\,\frac{x_{2}}{l_{2}%
}\right)  \right\}  ,\label{g12}%
\end{equation}
and%

\begin{align}
G_{2}(x_{1},x_{2},x_{3}) &  =-\frac{2\pi}{l_{1}l_{2}}x_{3}+\frac{1}{\pi}%
\frac{1}{l_{1}l_{2}}\sum_{m_{1}^{^{\prime}},m_{2}}\frac{\pi}{\sqrt{\left(
\frac{m_{1}}{l_{1}}\right)  ^{2}+\left(  \frac{m_{2}}{l_{2}}\right)  ^{2}}%
}\label{g4}\\
&  \times\exp\left(  -2\pi\sqrt{\left(  \frac{m_{1}}{l_{1}}\right)
^{2}+\left(  \frac{m_{2}}{l_{2}}\right)  ^{2}}|x_{3}|\right)  \prod_{i=1}%
^{2}\cos\left(  2\pi m_{i}\,\frac{x_{i}}{l_{i}}\right)  .\nonumber
\end{align}
First we obtain $G_{1}$ in a closed form as follows. We may employ the identity
from Eq.(\ref{iden1}) to obtain%

\begin{align}
G_{1}(x_{2},x_{3})  &  =-\frac{1}{l_{1}}\ln\left[  \cosh\left(  2\pi
\frac{x_{3}}{l_{2}}\right)  -\cos\left(  2\pi\frac{x_{2}}{l_{2}}\right)
\right] \label{gx3}\\
&  -\frac{\ln\left(  2\right)  }{l_{1}}+2\pi\frac{|x_{3}|}{l_{1}l_{2}}.\nonumber
\end{align}
As discussed in the appendix A, $G_{1}$ has a logarithmic divergence when
$x_{2}/l_{2}$ and $x_{3}/l_{2}$ tend to zero. As we will see soon, a similar
logarithmic divergence with opposite sign arises from the term $G_{2}$. These
two divergences cancel each other to give a finite contribution to
$G_{\text{slab}}$ towards small values of $x_2$ and $x_3$.

We consider the case of $G_{2}$ from Eq.(\ref{g4}). Applying the Poisson
summation rule \cite{arnold}, the sum over $m_{2}$ in Eq.(\ref{g4}) may be
transformed to a sum involving Bessel functions of the second
kind\cite{arnold}:%

\begin{align}
\frac{1}{\left\vert \delta\right\vert }\sum_{m}\pi\frac{\exp\left(
-|z|\sqrt{\alpha^{2}+\left(  \frac{2\pi}{\delta}m\right)  ^{2}}\right)  }%
{\sqrt{\alpha^{2}+\left(  \frac{2\pi}{\delta}m\right)  ^{2}}}  &  \exp\left(
2\pi i~m\frac{x}{\delta}\right) \\
&  =\sum_{m}K_{0}\left(  \alpha\sqrt{z^{2}+(x+\delta m)^{2}}\right)
.\nonumber
\end{align}
Identifying%

\begin{equation}
\delta=l_{2},~z=x_{3},~\alpha=2\pi\frac{\left\vert m_{1}\right\vert }{l_{1}%
}\text{ and}~x=x_{2},
\end{equation}
we can write%

\begin{align}
G_{2}(x_{1},x_{2},x_{3})  &  =\frac{2}{l_{1}}\sum_{m_{1}^{\prime},m_{2}}%
K_{0}\left(  2\pi\frac{\left\vert m_{1}\right\vert }{l_{1}}\sqrt{(x_{2}%
+m_{2}l_{2})^{2}+x_{3}^{2}}\right) \label{g44}\\
&  \times\cos\left(  2\pi m_{1}\frac{x_{1}}{l_{1}}\right)  -\frac{2\pi}%
{l_{1}l_{2}}|x_{3}|.\nonumber
\end{align}
The sum in Eq.(\ref{g44}) may be expressed in two parts as%

\begin{align}
G_{2}(x_{1},x_{2},x_{3}) &  =\frac{2}{l_{1}}\sum_{m_{1}^{\prime},m_{2}%
^{\prime}}K_{0}\left(  2\pi\frac{\left\vert m_{1}\right\vert }{l_{1}}%
\sqrt{(x_{2}+m_{2}l_{2})^{2}+x_{3}^{2}}\right)  \cos\left(  2\pi m_{1}%
\frac{x_{1}}{l_{1}}\right)  \label{g444}\\
&  +\frac{2}{l_{1}}\sum_{m_{1}^{\prime}}K_{0}\left(  2\pi\frac{\left\vert
m_{1}\right\vert }{l_{1}}\sqrt{x_{2}^{2}+x_{3}{}^{2}}\right)  \cos\left(  2\pi
m_{1}\frac{x_{1}}{l_{1}}\right)  -\frac{2\pi}{l_{1}l_{2}}|x_{3}|.\nonumber
\end{align}
We note that the first term in Eq.(\ref{g444}) has no convergence problem as
$x_{2}$ and $x_{3}$ are positive numbers and $l_{2} \geq l_{1}$. This term will
convergence even for the case when $0 \leq x_{2}$ and $x_{3}$ are zero. The
convergence of $G_{2}$ and thus that of $G_{\text{slab }}$and $G_{\text{3d}}$
depend upon the ratio%
\begin{equation}
\rho=\frac{\left(  x_{2}^{2}+x_{3}^{2}\right)  ^{1/2}}{l_{1}},
\label{rho}
\end{equation}
which appears in the second term on the rhs of Eq.(\ref{g444}). For
$\rho>0.1$, Eq.(\ref{g444}) will have a very good convergence. However, if
$x_{2}$ and $x_{3}$ are such that the condition $\rho>0.1$ is not satisfied
then we should transform Eq.(\ref{g444}) further. This can be done by
using the results derived in appendix B
where it is shown that
\begin{align}
f\left(  x_{1},x_{2},x_{3}\right)   &  =\frac{4}{l_{1}}\sum_{m_{1}=1}^{\infty
}\,K_{0}\left(  \frac{2\pi m_{1}}{l_{1}}\left(  x_{2}^{2}+x_{3}^{2}\right)
^{1/2}\right)  \cos\left(  \frac{2\pi m_{1}}{l_{1}}x_{1}\right)  \label{f}\\
&  =\frac{2}{l_{1}}\left\{  \gamma+\ln\left(  \frac{\left(  x_{2}^{2}%
+x_{3}^{2}\right)  ^{1/2}}{2l_{1}}\right)  \right\}  +\frac{1}{\sqrt{x_{1}%
^{2}+x_{2}^{2}+x_{3}^{2}}}\nonumber\\
&  +\frac{1}{l_{1}}\sum_{n_{1}=1}^{N-1}\left(  \frac{1}{\sqrt{\rho^{2}+\left(
n_{1}+x\right)  ^{2}}}+\frac{1}{\sqrt{\rho^{2}+\left(  n_{1}-x\right)  ^{2}}%
}\right)  \nonumber\\
&  -\frac{2\gamma}{l_{1}}-\frac{\left\{  \psi(N+x)+\psi(N-x)\right\}  }{l_{1}%
}\nonumber\\
&  +\frac{1}{l_{1}}\sum_{l=1}^{\infty}\binom{-1/2}{l}\rho^{2l}\left(
\zeta\left(  2l+1,N+x\right)  +\zeta\left(  2l+1,N-x\right)  \right)
,\nonumber
\end{align}
where $x=x_{1}/l_{1}$ and $\psi$ and $\zeta$ stand for digamma and Hurwitz
Zeta function respectively. $N\geq1$ is the smallest integer satisfying the
condition $N>\rho+x$. Thus we can choose $N=1$, as even for the worst case one
has $\rho=0.1$ and $x=0.5$ . However, for better convergence it is desirable
that one chooses $N$ such that $N>\rho+1$. 

We can now write the following
short algorithm to calculate $G_{\text{slab }}$. First we set our axis such
that $l_{3}\geq l_{2}\geq l_{1}.$ Next, using the periodic boundary
conditions, the separation between two
particles can always be reduced in such a way that the individual
components satify $0 \leq x_i < l_i$. Thus, the
values of $r_{i}=x_{i}/l_{i}$ lie
between $0$ and $1.$ From the inherent symmetry of the problem, energy
corresponding to eight different separations of $\left(  \frac{1\pm r_{1}}{2},\frac{1\pm
r_{2}}{2},\frac{1\pm r_{3}}{2}\right)  $ is the same. This essentially
means that we can concentrate our attention on only on those
separations between the particles which correspond to $0 \leq
r_{i}\leq 0.5.$ If some $r_{i}>0.5$, we can replace it
with $1-r_{i}$. Next, we look at the value of $r_{32}=r_{3}/l_{2}.$ If
$r_{32}>0.1$, we can combine Eq.(\ref{gx3}) with Eq.(\ref{g4}) to obtain the
following form for $G_{\text{slab}}$%

\begin{align}
G_{\text{slab}}(x_{1},x_{2},x_{3}) &  =-\frac{1}{l_{1}}\ln\left[  \cosh\left(
2\pi\frac{x_{3}}{l_{2}}\right)  -\cos\left(  2\pi\frac{x_{2}}{l_{2}}\right)
\right]\label{far}  \\
&  -\frac{\ln\left(  2\right)  }{l_{1}}+\frac{1}{\pi}\frac{1}{l_{1}l_{2}}%
\sum_{m_{1}^{^{\prime}},m_{2}}\frac{\pi}{\sqrt{\left(  \frac{m_{1}}{l_{1}%
}\right)  ^{2}+\left(  \frac{m_{2}}{l_{2}}\right)  ^{2}}}\nonumber\\
&  \times\exp\left(  -2\pi\sqrt{\left(  \frac{m_{1}}{l_{1}}\right)
^{2}+\left(  \frac{m_{2}}{l_{2}}\right)  ^{2}}|x_{3}|\right)  \prod_{i=1}%
^{2}\cos\left(  2\pi m_{i}\,\frac{x_{i}}{l_{i}}\right)  .\nonumber
\end{align}
However if $0 \leq r_{32}<0.1$, then we look at the value of $\rho$, which is
defined in Eq.(\ref{rho}). If $\rho>0.1$,
we should use the following form of $G_{\text{slab}}$ which is
obtained after combining Eqs (\ref{gx3}) and (\ref{g444}):%

\begin{align}
G_{\text{slab}}(x_{1},x_{2},x_{3}) &  =-\frac{1}{l_{1}}\ln\left[  \cosh\left(
2\pi\frac{x_{3}}{l_{2}}\right)  -\cos\left(  2\pi\frac{x_{2}}{l_{2}}\right)
\right]  \\
&  -\frac{\ln\left(  2\right)  }{l_{1}}+\frac{2}{l_{1}}\sum_{m_{1}^{^{\prime}%
},m_{2}^{^{\prime}}}K_{0}\left(  2\pi\frac{\left\vert m_{1}\right\vert }%
{l_{1}}\sqrt{(x_{2}+m_{2}l_{2})^{2}+x_{3}^{2}}\right)  \cos\left(  2\pi
m_{1}\frac{x_{1}}{l_{1}}\right)  \nonumber\\
&  +\frac{4}{l_{1}}\sum_{m_{1}=1}^{\infty}\,K_{0}\left(  \frac{2\pi m_{1}%
}{l_{1}}\left(  x_{2}^{2}+x_{3}^{2}\right)  ^{1/2}\right)  \cos\left(
\frac{2\pi m_{1}}{l_{1}}x_{1}\right)  .\nonumber
\end{align}
If $\rho<0.1$ then we use the identity in Eq.(\ref{f}) to write $G_{\text{slab}}$ as%
\begin{align}
G_{\text{slab}}(x_{1},x_{2},x_{3}) &  =-\frac{1}{l_{1}}\ln\left[  \cosh\left(
2\pi\frac{x_{3}}{l_{2}}\right)  -\cos\left(  2\pi\frac{x_{2}}{l_{2}}\right)
\right]  \label{fg1}\\
&  -\frac{\ln\left(  2\right)  }{l_{1}}+\frac{2}{l_{1}}\sum_{m_{1}^{^{\prime}%
},m_{2}^{^{\prime}}}K_{0}\left(  2\pi\frac{\left\vert m_{1}\right\vert }%
{l_{1}}\sqrt{(x_{2}+m_{2}l_{2})^{2}+x_{3}^{2}}\right)  \cos\left(  2\pi
m_{1}\frac{x_{1}}{l_{1}}\right)  \nonumber\\
&  +\frac{2}{l_{1}}\left\{  \gamma+\ln\left(  \frac{\left(  x_{2}^{2}%
+x_{3}^{2}\right)  ^{1/2}}{2l_{1}}\right)  \right\}  +\frac{1}{\sqrt{x_{1}%
^{2}+x_{2}^{2}+x_{3}^{2}}}\nonumber\\
&  +\frac{1}{l_{1}}\sum_{n_{1}=1}^{N-1}\left(  \frac{1}{\sqrt{\rho^{2}+\left(
n_{1}+x\right)  ^{2}}}+\frac{1}{\sqrt{\rho^{2}+\left(  n_{1}-x\right)  ^{2}}%
}\right)  \nonumber\\
&  -\frac{2\gamma}{l_{1}}-\frac{\left\{  \psi(N+x)+\psi(N-x)\right\}  }{l_{1}%
}\nonumber\\
&  \frac{1}{l_{1}}\sum_{l=1}^{\infty}\binom{-1/2}{l}\rho^{2l}\left(
\zeta\left(  2l+1,N+x\right)  +\zeta\left(  2l+1,N-x\right)  \right)
.\nonumber
\end{align}

Although Eq.(\ref{fg1}) is meant to be used only when $\rho<0.1$, the
equation is defined for all values of $\rho$ as long as $N$ is chosen
such that $N>\rho+1$. The series given in Eq.(\ref{fg1}) is valid when both
$x_{2}$ and $x_{3}$ are non zero. In this case, the argument of the first
logarithmic term on the rhs of Eq.(\ref{fg1}) is always greater than zero. However,
for very small values of $x_{2}$ and $x_{3}$ (say both less than
$\varepsilon=10^{-3}$) the first and the fourth terms diverge. In such
situation one should combine the diverging terms together using the function $L$ defined
in appendix B. 

We have thus shown how to compute $G_{\text{slab}}$ for all regions of
the unit cell. Similar results for the slab geometry have previously been
obtained in Refs. \onlinecite{arnold}, \onlinecite{liem} and
\onlinecite{niels1}. Results in Eqs.(\ref{far}) and (\ref{fg1}) correspond
respectively to "near" and "far" formulae derived by Arnold \textit{et
al}.\cite{arnold}. Also, it is an easy matter now to obtain expressions 
for $G_{\text{3d}}$ from Eq.(\ref{combine}). One can obtain the self
energy for a 3D system as,

\begin{align}
G_{\text{self}}^{\text{3d}} &  =\lim_{\left(  x_{1},x_{2},x_{3}\right)  \rightarrow\left(
0,0,0\right)  }\left(  G_{\text{3d}}(x_{1},x_{2},x_{3})-\frac{1}{\sqrt
{x_{1}^{2}+x_{2}^{2}+x_{3}^{2}}}\right)  \label{self3d}\\
&  =\frac{1}{\pi}\frac{l_{3}}{l_{1}l_{2}}\sum_{m_{1},m_{2}}^{\prime}\frac{\pi
}{\gamma_{30}(\{m\})}\frac{\exp\left(  -\pi\gamma_{30}(\{m\})\right)  }%
{\sinh(\pi\gamma_{30}(\{m\}))}\nonumber\\
&  +\frac{2}{l_{1}}\sum_{m_{1}^{^{\prime}},m_{2}^{^{\prime}}}K_{0}\left(
2\pi\left\vert m_{1}m_{2}\right\vert \frac{l_{2}}{l_{1}}\right)  +\frac{l_{3}%
}{l_{1}l_{2}}\frac{\pi}{3}-\frac{2}{l_{1}}\ln\left(  \frac{4\pi l_{1}}{l_{2}%
}\right)  +\frac{2\gamma}{l_{1}},\nonumber
\end{align}
where for $G_{\text{3d}}$ we use Eqs. (\ref{combine}) and (\ref{fg1}).

\section{Madelung Constants}

Using the formulas developed above, it is an easy matter to obtain expressions
for the Madelung constants of NaCl and CsCl. A simple structural analysis of
CsCl easily leads to expression,%

\begin{equation}
M_{\text{CsCl}}=G_{\text{3d}}\left(  \frac{1}{2},\frac{1}{2},\frac{1}{2}\right)
-G_{\text{self}}^{\text{3d}},
\end{equation}
and similarly for NaCl we see%

\begin{equation}
M_{\text{NaCl}}=\frac{1}{2}\left[  G_{\text{3d}}\left(  \frac{1}{2},\frac{1}{2},\frac
{1}{2}\right)  +3G_{\text{3d}}\left(  0,0,\frac{1}{2}\right)  -3G_{\text{3d}}\left(  \frac{1}%
{2},0,\frac{1}{2}\right)  -G_{\text{self}}^{\text{3d}}\right]  ,
\end{equation}
where Eq.(\ref{combine}) can be used for $G_{\text{3d}}(x_{1},x_{2},x_{3})$ with all $l_{i}$'s
set equal to one. From the above equations we obtain the following expressions
for Madelung constants of CsCl and NaCl:%

\begin{align}
M_{\text{CsCl}}  &  =-\frac{1}{\pi}\sum_{m_{1},m_{2}}^{\prime}\frac{\pi}%
{\sqrt{m_{1}^{2}+m_{2}^{2}}}\frac{\left[  \exp\left(  -\pi\sqrt{m_{1}%
^{2}+m_{2}^{2}}\right)  -\left(  -1\right)  ^{m_{1}+m_{2}}\right]  }{\sinh
(\pi\sqrt{m_{1}^{2}+m_{2}^{2}})}\\
&  -2\left(  \sum_{m_{1}^{^{\prime}},m_{2}^{^{\prime}}}K_{0}\left(
2\pi\left\vert m_{1}m_{2}\right\vert \right)  -\ln\left(  4\pi\right)
+\gamma-\pi\right)  ,\nonumber
\end{align}
and%

\begin{align}
2~M_{\text{NaCl}}  &  =-\frac{1}{\pi}\sum_{m_{1},m_{2}}^{\prime}\frac{\pi
}{\sqrt{m_{1}^{2}+m_{2}^{2}}}\frac{\left(  \exp\left[  -\pi\sqrt{m_{1}%
^{2}+m_{2}^{2}}\right]  -\left(  -1\right)  ^{m_{1}+m_{2}}-3+3\left(
-1\right)  ^{m_{1}}\right)  }{\sinh(\pi\sqrt{m_{1}^{2}+m_{2}^{2}})}\\
&  -2\left(  \sum_{m_{1}^{^{\prime}},m_{2}^{^{\prime}}}K_{0}\left(
2\pi\left\vert m_{1}m_{2}\right\vert \right)  -\ln\left(  4\pi\right)
+\gamma-\pi\right)  .\nonumber
\end{align}
Restricting the sum over $m_{1}$ and $m_{2}$ between $-4$ and $+4$, a simple
calculation on Mathematica gives a $M_{\text{CsCl}}$ value correct up to
$10^{-8}$ and $M_{\text{NaCl}}$ value correct up to $10^{-6}$. In addition we
also obtain a simple relationship between the two Madelung constants:%

\begin{equation}
2~M_{\text{NaCl}}=M_{\text{CsCl}}+6\sum_{m_{1},m_{2}}\frac{\csc\left(
\pi\sqrt{\left(  2m_{1}+1\right)  ^{2}+m_{2}^{2}}\right)  }{\sqrt{\left(
2m_{1}+1\right)  ^{2}+m_{2}^{2}}}.
\end{equation}
This interesting relationship was first established by Hautot \cite{hautot} in
seventies using Hankel integrals and Schloimilch series.

\section{Conclusion}

Complete expressions for Coulomb sum for a rectangular cell in 2D and an
orthorhombic cell in 3D were derived. We also obtained expressions
for the self energies. The expressions obtained provide convergence in all
parts of the unit cell. Considerable simplification has been achieved over
Sperb's work\cite{sperb2} in terms of deriving the equations. 
The proposed formula for the potential energy when
the two charges are very close, differs from that of Sperb. In particular,
when the charges are close together, Sperb's\cite{sperb2} formula has a triple sum
(Eq. 2.4 and 2.7). In our
expression, we have at most a double sum. Similar results for 3D case
have previously been obtained by Strebel using a rather involved
procedure \cite{strebel}. 
Our results do not require any
convergence parameter like that used in Ewald sums, neither do our formulas
involve any complementary error functions. These error functions in an Ewald
sum are a source of loss of precision when calculating Madelung constants to
higher accuracies.

In retrospect, we see that these results could be derived in another way by
starting off with the Green function expression for the $2D+h$ slab geometry
system and then adding the ELC term which takes into account the rest
of the layers. This way we will get only the first two terms of
Eq.(\ref{combine}). The third term is then obtained by
adding a term proportional to $M_3^2$ from outside, where $M_3$ stands for 
the component of the total dipole moment along the $x_3$ direction. In
the present work this dipole term arises naturally, as shown in Eq. (\ref{dipole}). 
This dipole term has been discussed by Smith \cite{smith}. Thus this
slabwise summation plus a term dipole term added from outside,apart
from an unimportant constant, leads to the same expression as in
Eq.(\ref{combine}). 
Thus, our Eqs.(\ref{combine}) and (\ref{dipole}) can be viewed as 
an alternative derivation of Eq.(4) in Ref. \onlinecite{arnold2}.

An advantage of the method developed here is that one can achieve better
time scaling in a simulation. Using the expressions presented in this paper,
the time to calculate forces and energy for a 3D system in a computer
simulation scales as $N^{2}$, where $N$ is the number of charges present in
the unit cell. However, one can achieve $N^{5/3}\ln(N)^2$ scaling after a little
modification in the expressions presented here. This is the same scaling as
achieved by Arnold \textit{et al.}\cite{arnold} for $2D+h$ system. The scaling
remains the same for the two cases because the electrostatic correction term
can be computed linearly if we remove the contribution of the first two
closest layers enclosing the unit cell in a given direction as opposed to
removing the contribution of just one layer as done by Arnold \textit{et al.}
\cite{arnold2} and in this paper. Also the results presented here can be
generalized to a rhombic cell in 2D and a triclinic cell in
3D\cite{paper3}

Our proposed expressions can be applied to calculation of Madelung constants
in 3D. Results obtained for the Madelung constants of CsCl and NaCl match with
those in the literature.

In conclusion we have provided a very simple derivation of complicated
results previously obtained by many authors using different, sometimes
complicated, techniques \cite{sperb2, lekner1, arnold, liem, niels1}.

\acknowledgments{ I am thankful to Dr. Y. Y. Goldschmidt for useful discussions. I also
thank Philip C. Tillman and Mahesh Bandi for suggesting improvements in the paper.}

\appendix

\section{Logarithmic Divergence}

Consider the function
\begin{equation}
L(x,y)=\ln\left[  \cosh y-\cos x\right]  -\ln\left[  \frac{y^{2}+x^{2}}%
{2}\right]  . \label{ap1}%
\end{equation}
We want to examine the limiting value of $L$ as $x$ and $y$ tend to zero. For
this reason we expand the argument of the first logarithmic term in Eq.
(\ref{ap1}),%
\begin{align}
\cosh y-\cos x  &  =\left(  \frac{y^{2}+x^{2}}{2!}\right)  +\left(
\frac{y^{4}-x^{4}}{4!}\right)  +\left(  \frac{y^{6}+x^{6}}{6!}\right)
\label{ap2}\\
&  +\left(  \frac{y^{8}-x^{8}}{8!}\right)  +\text{O}\left[  x^{10}%
,y^{10}\right]  .\nonumber
\end{align}
Factoring out the first term on the right hand side, Eq.(\ref{ap2}) can be
written as%
\begin{align}
\cosh y-\cos x  &  =\left(  \frac{y^{2}+x^{2}}{2!}\right)  \left\{
1+\frac{2!}{4!}\left(  y^{2}-x^{2}\right)  +\frac{2!}{6!}\left(  y^{4}%
-x^{2}y^{2}+x^{4}\right)  \right. \label{ap3}\\
&  \left.  +\frac{2!}{8!}\left(  y^{4}+x^{4}\right)  \left(  y^{2}%
-x^{2}\right)  +\text{O}\left[  x^{8},y^{8}\right]  \right\}  .\nonumber
\end{align}
Thus $L$ can be written as%
\begin{align}
L(x,y)  &  =\ln\left\{  1+\frac{2!}{4!}\left(  y^{2}-x^{2}\right)  +\frac
{2!}{6!}\left(  y^{4}-x^{2}y^{2}+x^{4}\right)  \right. \label{ap4}\\
&  \left.  +\frac{2!}{8!}\left(  y^{4}+x^{4}\right)  \left(  y^{2}%
-x^{2}\right)  +\text{O}\left[  x^{8},y^{8}\right]  \right\}  .\nonumber
\end{align}
Using the results from Eqs. (\ref{ap1}) and (\ref{ap4}) in Eq.(\ref{gx3}), we
see that for small values of $x_{2}/l_{2}$ and $x_{3}/l_{2}$, $G_{1}$ can be
written as
\begin{align}
G_{1}(x_{2},x_{3})  &  =-\frac{1}{l_{1}}\ln\left[  2\pi^{2}\frac{\left(
x_{2}^{2}+x_{3}^{2}\right)  }{l_{2}^{2}}\right]  -\frac{\ln\left(  2\right)
}{l_{1}}+2\pi\frac{|x_{3}|}{l_{1}l_{2}}\label{ap5}\\
&  -\frac{1}{l_{1}}L\left(\frac{x_2}{l_2},\frac{x_3}{l_2}\right),\nonumber
\end{align}
which clearly shows a logarithmic divergence as $x_{2}/l_{2}$ and $x_{3}%
/l_{2}$ tend to zero.

\subsection{Line Charge}

We commence with the identity \cite{gradshteyn},%

\begin{align}
f\left(  x_{1},x_{2},x_{3}\right)   &  =\frac{4}{l_{1}}\sum_{m_{1}=1}^{\infty
}\,K_{0}\left(  \frac{2\pi m_{1}}{l_{1}}\left(  x_{2}^{2}+x_{3}^{2}\right)
^{1/2}\right)  \cos\left(  \frac{2\pi m_{1}}{l_{1}}x_{1}\right) \label{iden2}%
\\
&  =\frac{2}{l_{1}}\left\{  \gamma+\ln\left(  \frac{\left(  x_{2}^{2}%
+x_{3}^{2}\right)  ^{1/2}}{2l_{1}}\right)  \right\}  +\frac{1}{\sqrt{x_{1}%
^{2}+x_{2}^{2}+x_{3}^{2}}}+S\left(  x_{1},x_{2},x_{3}\right)  ,\nonumber
\end{align}
where%
\begin{align}
S\left(  x_{1},x_{2},x_{3}\right)  =\sum_{n=1}^{\infty}  &  \left(  \frac
{1}{\sqrt{x_{2}^{2}+x_{3}^{2}+\left(  nl_{1}-x_{1}\right)  ^{2}}}\right. \\
&  +\left.  \frac{1}{\sqrt{x_{2}^{2}+x_{3}^{2}+\left(  nl_{1}+x_{1}\right)
^{2}}}-\frac{2}{n}\right)  .
\end{align}
We can further transform the identity in Eq.(\ref{iden2}) along the lines
worked out by Strebel \cite{strebel} and Arnold {\it et al.}\cite{arnold}.
Let us look at%

\begin{align}
h(\rho,x_{1}) &  =\frac{1}{l_{1}}\sum_{n=N}^{\infty}\left(  \frac{1}%
{\sqrt{\rho^{2}+\left(  n+\frac{x_{1}}{l_{1}}\right)  ^{2}}}-\frac{1}%
{n}\right)  \\
&  =\frac{1}{l_{1}}\sum_{n=1}^{\infty}\left(  \frac{1}{\sqrt{\rho^{2}+\left(
n+N-1+\frac{x_{1}}{l_{1}}\right)  ^{2}}}-\frac{1}{n+N-1}\right)  \\
&  =\frac{1}{l_{1}}\sum_{n=1}^{\infty}\left(  \frac{1}{\sqrt{\rho^{2}+\left(
n+y\right)  ^{2}}}-\frac{1}{n}\right)  +\frac{1}{l_{1}}\sum_{n=1}^{N}\frac
{1}{n},
\end{align}
where $N\geq1,$ $y=N-1+x_{1}/l_{1}$ and%
\begin{equation}
\rho=\frac{\left(  x_{2}^{2}+x_{3}^{2}\right)  ^{1/2}}{l_{1}},~x=\frac{x_{1}%
}{l_{1}}.
\end{equation}
Assuming $\rho<|1+y|$, the Binomial expansion of the first term in the above
equation gives%

\begin{align}
\frac{1}{\sqrt{\rho^{2}+\left(  n+y\right)  ^{2}}}  &  =\sum_{p=0}^{\infty
}\binom{-1/2}{p}\rho^{2p}\frac{1}{\left\vert n+y\right\vert ^{2p+1}}\\
&  =\sum_{p=1}^{\infty}\binom{-1/2}{p}\rho^{2p}\frac{1}{\left\vert
n+y\right\vert ^{2p+1}}+\frac{1}{\left\vert n+y\right\vert },\nonumber
\end{align}
where $\binom{-1/2}{p}$ stands for the Binomial coefficient. We can take the
sum over $n$ inside and obtain%

\begin{align}
h\left(  \rho,x_{1}\right)   &  =\frac{1}{l_{1}}\sum_{p=1}^{\infty}%
\binom{-1/2}{p}\rho^{2p}\sum_{n=1}^{\infty}\frac{1}{\left\vert n+y\right\vert
^{2p+1}}\label{hr1}\\
&  +\frac{1}{l_{1}}\sum_{n=1}^{\infty}\left(  \frac{1}{\left\vert
n+y\right\vert }-\frac{1}{n}\right)  +\frac{1}{l_{1}}\sum_{n=1}^{N}\frac{1}%
{n}.
\end{align}
Now, using the definition of the Hurwitz Zeta function,%

\begin{equation}
\zeta\left(  l,y\right)  =\sum_{k=0}^{\infty}\frac{1}{\left(  k+y\right)
^{l}},
\end{equation}
we obtain%

\begin{equation}
\sum_{n=1}^{\infty}\frac{1}{\left\vert n+y\right\vert ^{2p+1}}=\zeta\left(
2p+1,1+y\right)  . \label{zeta}%
\end{equation}
Also the second sum in Eq.(\ref{hr1}) is easy to obtain. By the definition of
the digamma function $\psi$ we have%

\begin{equation}
\sum_{n=1}^{\infty}\left(  \frac{1}{\left\vert n+y\right\vert }-\frac{1}%
{n}\right)  =-\gamma-\psi\left(  1+y\right)  . \label{digamma}%
\end{equation}
Thus $h\left(  \rho,x_{1}\right)  $ can be written as%
\begin{equation}
h\left(  \rho,x_{1}\right)  =-\frac{\gamma}{l_{1}}-\frac{\psi\left(
1+y\right)  }{l_{1}}+\frac{1}{l_{1}}\sum_{l=1}^{\infty}\binom{-1/2}{l}%
\rho^{2l}\zeta\left(  2l+1,1+y\right)  +\frac{1}{l_{1}}\sum_{n=1}^{N}\frac
{1}{n}%
\end{equation}
Using Eqs. (\ref{hr1}), (\ref{zeta}) and (\ref{digamma}) we obtain,%

\begin{align}
S\left(  x_{1},x_{2},x_{3}\right)   &  =\frac{1}{l_{1}}\sum_{n=1}^{N-1}\left(
\frac{1}{\sqrt{\rho^{2}+\left(  n+x_{1}\right)  ^{2}}}+\frac{1}{\sqrt{\rho
^{2}+\left(  n-x_{1}\right)  ^{2}}}-\frac{2}{n}\right)  \label{sx}\\
&  +\frac{1}{l_{1}}\sum_{n=N}^{\infty}\left(  \frac{1}{\sqrt{\rho^{2}+\left(
n+x_{1}\right)  ^{2}}}+\frac{1}{\sqrt{\rho^{2}+\left(  n-x_{1}\right)  ^{2}}%
}-\frac{2}{n}\right)  \nonumber\\
&  =\frac{1}{l_{1}}\sum_{n=1}^{N-1}\left(  \frac{1}{\sqrt{\rho^{2}+\left(
n+x_{1}\right)  ^{2}}}+\frac{1}{\sqrt{\rho^{2}+\left(  n-x_{1}\right)  ^{2}}%
}\right)  +\frac{1}{\sqrt{x_{1}^{2}+x_{2}^{2}+x_{3}^{2}}}\nonumber\\
&  -\frac{2\gamma}{l_{1}}-\frac{\psi(N+x_{1})+\psi(N-x_{1})}{l_{1}}\nonumber\\
&  +\frac{1}{l_{1}}\sum_{l=1}^{\infty}\binom{-1/2}{l}\rho^{2l}\left(
\zeta\left(  2l+1,N+x\right)  +\zeta\left(  2l+1,N-x\right)  \right)
,\nonumber
\end{align}
Note that for Eq.(\ref{sx}) to be valid, the condition is that $\rho<|1+y|$,
where $y=N-1\pm x_{1}/l_{1}$. Keeping in mind that $x_{1}\geq0$ we get
$\rho<|N\pm x_{1}/l_{1}|$, which will be satisfied if $N>\rho+x.$ Combining
Eq.(\ref{iden2}) and Eq.(\ref{sx}) gives us the result that we set out to prove.

\end{document}